# Crowding Effects during DNA Translocation in Nanopipettes

*Rand A. Al-Waqfi$^{\$,£}$, Cengiz Khan$^{\$}$, Oliver J. Irving$^{\$}$, Lauren Matthews$^{\$,\&}$, Tim Albrecht$^{\$}$*

$^{\$}$ University of Birmingham, School of Chemistry, Edgbaston Campus, Birmingham B15 2TT, United Kingdom

$^{£}$ Department of Medicinal Chemistry and Pharmacognosy, Faculty of Pharmacy, Jordan University of Science and Technology, P.O. Box 3030, Irbid 22110, Jordan

$^{\&}$ Federal Institute for Materials Research and Testing, Department 6, Unter den Eichen 87, 12205 Berlin, Germany

Quartz nanopipettes are an important emerging class of electric single-molecule sensors for DNA, proteins, their complexes as well as other biomolecular targets. However, in comparison to other resistive pulse sensors, nanopipettes constitute a highly asymmetric environment and the transport of ions and biopolymers can become strongly direction-dependent. For double-stranded DNA, this can include the characteristic translocation time and its tertiary structure, but as we show here, nanoconfinement can not only lead to unexplored features in the transport characteristics of the sensor, but also unlock new capabilities for biophysical and bioanalytical studies at the single-molecule level. To this end, we show how the accummulation of DNA inside the nanochannel leads to crowding effects, and in some cases reversible blocking of DNA entry, and provide a detailed analysis based on a range of different DNA samples and experimental conditions. Moreover, using biotin-functionalised DNA and streptavidin-modified gold nanoparticles as target, we demonstrate in a proof-of-concept study how the crowding effect, and the resulting increased residence time in nanochannel, can be exploited in a new analytical paradigm, involving DNA injection into the nanochannel, incubation with the nanoparticle target and analysis of the complex by reverse translocation. We thereby integrate elements of sample processing and detection into the nanopipette, as an important conceptual advance, and make a case for the wider applicability of this device concept.

Resistive-pulse sensors are an important class of single-molecule sensors and broadly fall into three classes, namely biological, solid-state or silicon chip-based nanopores, and nanopipettes.[1-10] They share certain common features, for example that typically a single channel connects two liquid compartments, which are otherwise separated through a highly insulating membrane. The transport of individual biomolecular analytes through the channel ("translocation") typically alters the resistance of the channel temporarily, resulting in a measurable, transient change in the electric current through the system.[1]

In biological nanopore sensors, the channel is often constituted by a pore forming protein, such as α-hemolysin, embedded in a lipid bilayer membrane. Very small pores can be fabricated with high precision in this way, which is why such pores are being used for DNA and, more recently, explored towards protein sequencing.[11] In chip-based nanopore sensors, pores are formed by electron or ion beam milling into a thin, solid-state membrane, for example made of silicon nitride, graphene or another 2D material, or by using strong, localized electric fields.[12, 13] More detailed reviews of these two classes can be found elsewhere.[13-15] Notably, however, in both of these, the transport of ions or biomolecular analytes between the bulk of the solution and the pore channel is comparable on both sides of the membrane. Accordingly, "capture and recapture" experiments, where an analyte is first translocated in one direction and then recaptured by a fast bias reversal, yielded similar results for both translocation directions, for example in terms of the observed average dwell times for double-stranded (ds) DNA and nanoparticles.[16-18]

On the other hand, nanopipette-based sensors are fabricated from (quartz) capillaries using a pipette puller, i.e. involving a combination of localized laser-induced heating and mechanical pulling. The pore channel is usually narrowest at the tip of the pipette, with inner diameters typically in the range of 5-20 nm, and then gradually widens up to the diameter of the capillary.[4,

[19] Hence, opening angles smaller than 10° and taper lengths of several mm are typical, while the sensing region is still confined to the narrowest part of the channel (say, the first 50 nm taken from the end of the tip).[20] As a result, compared to the outside of the pipette, transport on the inside is geometrically restricted, the electric field decays more gradually from the tip and surface effects more prominent[4]. In the past, this has been exploited for the localisation and controlled delivery of both single-stranded (ss) and dsDNA[21, 22] and more recently sparked renewed interest in understanding and exploiting the said asymmetry effects in the translocation of linear and functionalized dsDNA.[23-26]

Here, we provide a systematic analysis of dsDNA translocation in both directions, i.e. from the outside to the inside ("in") and the inside to the outside of the pipette ("out"), for a range of different DNA lengths from 4 kbp to 48.5 kbp and at different bias voltages. Importantly, and in contrast to previous literature, nanopipettes were loaded with 100s of dsDNA molecules for extended periods of time before initiating reverse translocation. This allows for the investigation of distinctly new features of the nanopipette sensor, such as crowding effects and the emergence of a new, distinct "translocation state" of the nanopipette once the number of translocated DNA molecules passes a certain DNA length-dependent threshold value. This new state is characterised by an increased translocation time and broadening of the translocation time distribution, however without a significant change in either pore conductance or electric noise. In the case of 10 kbp and 48.5 kbp λ-DNA, after continued operation translocation events were eventually no longer detected, while the pore conductance decreased (albeit not to zero) and the electrical noise increased. This effect was found to be reversible. Finally, we show that the temporary storage of functionalised DNA in the nanochannel can be exploited for incubation, target capture and read-out by reverse translocation. This is demonstrated in a proof-of-concept study for nanometer-sized, streptavidin-modified Au particles, but clearly opens up new avenues for integrated sample processing and read-out for other target analytes, such as biomarker proteins or RNA.

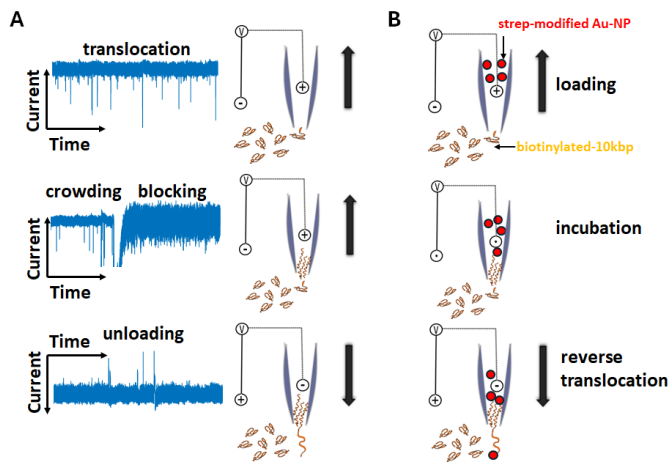

Figure 1. Illustration of two experimental paradigms described in this work. A) Translocation of long kbp DNA into the nanopipette (top), crowding of the nanochannel and eventually blocking of DNA translocation (centre) and, lastly, unloading of the nanopipette upon bias reversal (bottom). B) Capture of streptavidin-modified nanoparticles with functionalised (biotinylated) dsDNA after loading, incubation and reverse translocation for detection and analysis.

RESULTS AND DISCUSSION

As a first step, we investigated the translocation characteristics of 4 kbp DNA in 4 M LiCl, and in particular their dependence on the direction of transport into or out of the nanopipette. The results for the first four steps in the bias sequence, namely +0.8 V/-0.8 V/+0.6 V/-0.6 V, are summarised in figure 2 ($c_{DNA,out}$ = 300 pM, initially $c_{DNA,in}$ = 0). Recording of the current-time data, event detection and analysis are described in the Methods section. Example traces are shown in section 3 of the SI.

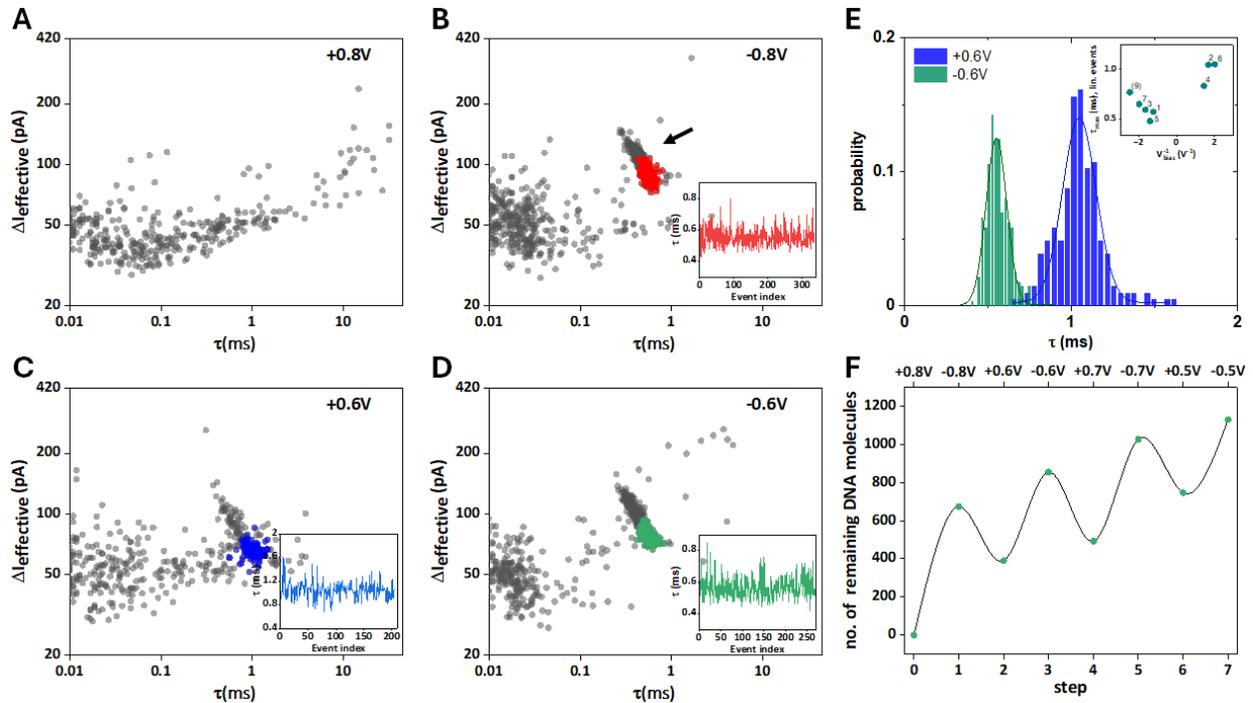

Figure 2: Translocation of 4 kbp DNA. A-D) Scatter plots of event magnitude vs. duration for steps 0-3 ($V_{bias}$ as indicated). Step 0 shows no DNA translocation, as there is no DNA present inside the nanopipette initially. Event clusters occur in subsequent steps (see arrow in panel B), linear events are colour-coded, as discussed in the main text. Insets: event duration vs. event index (linear events only). E) Translocation time distributions for linear events only, step 2 (blue) and step 3 (green), including normal distribution fits. Inset: a plot of $\tau_{max}$ vs. $V_{bias}^{-1}$ appears to be linear (steps 0-7 and 9), consistent with electrophoretically driven transport. The signal-to-noise ratio in step 8 ($V_{bias}$ = +0.4 V) was low and this dataset was therefore excluded from further analysis F) Number of DNA molecules inside the nanopipette, based on translocation statistics for steps 0-7.

Panels A-D show scatter plots of the effective event current magnitude, $\Delta I_{eff}$, vs. the dwell time $\tau$. For $V_{bias}$ = +0.8 V (panel A, step 0), given the bias polarity, no DNA translocation is expected, as initially $c_{DNA,in}$ = 0, and hence no DNA event cluster is observed.[5, 27, 28] Subsequently, $V_{bias}$ is switched to -0.8 V (B, step 1) and DNA translocates from the outside to the inside of the nanopipette. Accordingly, a distinct cluster of DNA translocation events appears at approximately 1 ms dwell time and an event magnitude of 100 pA (as indicated by the arrow; 673 events, average translocation frequency $f_1$ = 0.75 s$^{-1}$). Within this cluster, events at the top left are due to the translocation of folded DNA (and are thus shorter and more intense), while events at the bottom right more likely result from the translocation of DNA in a linear configuration (highlighted in red; cf. SI, section 4).[5,20] We formally identified 370 events as linear and 303 events as folded (linear/folded ratio = 0.55), which is in line with previous literature for nanopore of similar size.[19, 29] A plot of $\tau$ vs. the event index for those linear events shows no apparent correlation (inset), as expected for a stochastic process of this kind.

When $V_{bias}$ is changed to +0.6 V (C, step 2), some of the DNA that had previously entered the pipette in step 1 now translocates from the inside of the pipette to the outside (283 DNA events detected, $f_2$ = 0.3 s$^{-1}$). Linear events in the event cluster are highlighted in blue (linear/folded ratio = 0.77) and for those, there was again no apparent correlation between $\tau$ and event index. The series continues with $V_{bias}$ = -0.6 V (D, step 3), where 464 DNA events were detected (linear events in green, $f_3$ = 0.5 s$^{-1}$, linear/folded ratio = 0.59), and subsequently $V_{bias}$ = ± 0.7 V (steps 4 + 5) and $V_{bias}$ = ± 0.5 V (steps 6 + 7), see further details in the SI, section 8. Note that each $V_{bias}$ value is maintained for 1000s, so the average residence time of the DNA inside the nanopipette is longer than in experiments where the DNA is translocated and then rapidly recaptured.[24-26] A few observations are appropriate at this stage. Firstly, there was a moderate effect of the translocation direction on the linear/folded ratio, which was on average approximately 30% higher for translocation from the inside to the outside of the pipette. Secondly, the translocation time distribution was shifted to longer times but had a similar relative standard deviation, $\sigma/\tau_{max} \approx 0.1$, cf. panel E. This is further illustrated by a plot of the most probable translocation time $\tau_{max}$ vs. $V_{bias}^{-1}$, which shows a consistent upward shift of the positive branch, relative to the negative one, panel E inset. Furthermore, a linear correlation between $\tau_{max}$ and $V_{bias}^{-1}$ is consistent with electrophoretically driven transport of DNA.[1] Thirdly, the translocation frequencies f for the unloading steps 2, 4 and 6 are similar in magnitude, compared to the loading steps 1, 3, 5 and 7, even though the average concentrations of DNA on the outside and the inside of the pipette are expected to be very different. Taking steps 1 + 2 as an example, while $c_{DNA,out}$ = 300 pM, $c_{DNA,in}$ is given by the number of DNA molecules that translocated into the pipette in step 1 (673) and volume of electrolyte solution inside the pipette (~ 7 μL). This yields approximately, $c_{DNA,in} \approx 0.2$ fM. Ignoring differences in capture geometry and electric field distribution on the inside and the outside of the pipette for the time being, in electrophoretically dominated transport f ∝ $c_{DNA}$ and therefore the expected translocation frequency for the unloading steps should be about $10^6$-$10^7$ times smaller than for the loading steps. This is not the case, which suggests that the solution on the inside of the pipette, namely in the confined region, is not well mixed and that the local DNA concentration close to the nanopipette tip is increased.

Since the number of translocated DNA molecules is known for each step, the number of DNA molecules remaining inside the pipette can be estimated. The result of this analysis is depicted in panel F, which apart from oscillations due to consecutive loading/unloading cycles shows a steady increase over the course of the experiment. As shown below, this is qualitatively different for the longer DNA samples studied here.

In the case of 7 kbp DNA, a similar experimental design was followed, albeit with a bias sequence of ±0.5 V; ±0.8 V; ±0.6 V; ±0.7 V and a final step at +0.5 V ($c_{DNA,out}$ = 600 pM), fig. 3. Accordingly, no DNA translocation events are detected in step 0, since $c_{DNA,in}$ = 0 (panel A, $V_{bias}$ = +0.5 V). In step 1 ($V_{bias}$ = -0.5 V, panel B), DNA translocates into the nanopipette and an event cluster approximately 1 ms duration and 100 pA event magnitude is detected. Again, translocation events classified as linear are highlighted in red, the τ vs. event index plot is shown in the inset. No apparent correlation between the latter two parameters was observed for this subset of events. Interestingly, however, in the scatter plot there appears to be a tail of events with longer durations and magnitudes similar to linear DNA events, which was absent in the data obtained for 4 kbp DNA, cf. fig. 2 B.

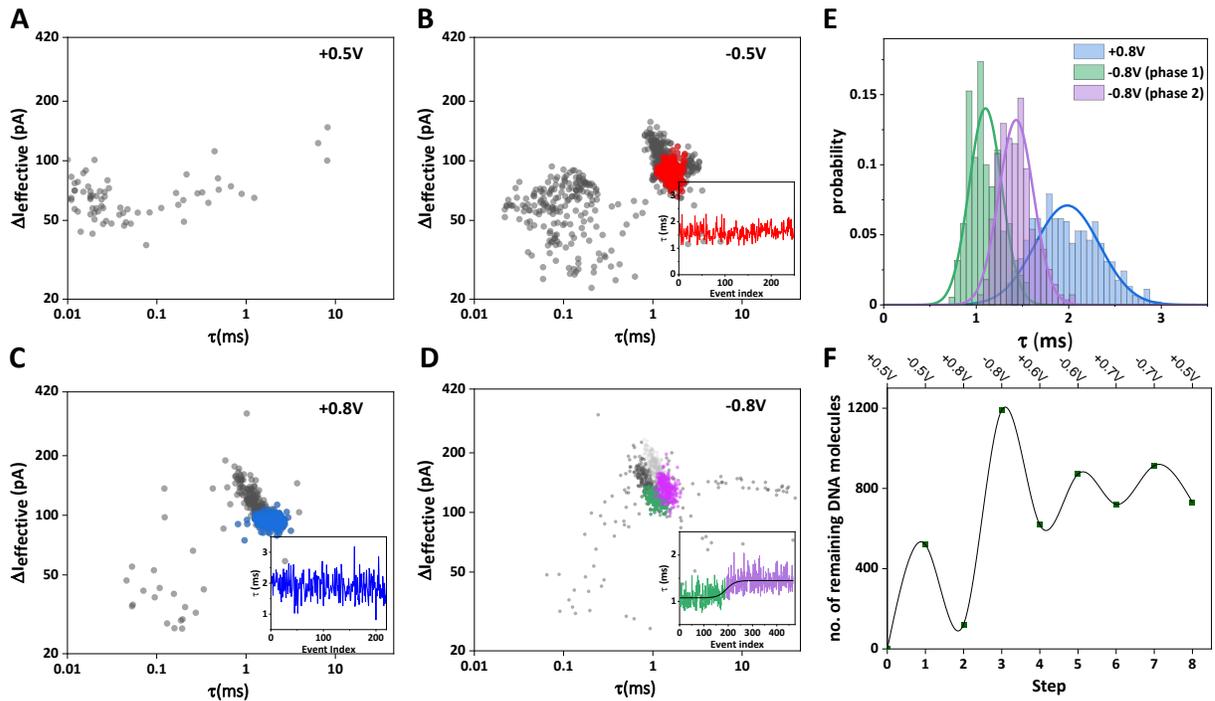

Figure 3. Translocation of 7 kbp DNA. A-D) Scatter plots of $\Delta I_{eff}$ vs. τ for steps 0-3 ($V_{bias}$ as indicated). Step 0 shows no DNA translocation, as there is initially no DNA present inside the nanopipette. Event clusters occur in subsequent steps, linear events are colour-coded, as discussed in the main text. Insets: τ vs. event index (linear events only). E) τ distributions for linear events only, step 2 (blue) and step 3, phase 1 (green) and phase 2 (magenta), including normal distribution fits. F) Number of DNA molecules inside the nanopipette, based on translocation statistics, for each step.

Unloading of the pipette occurs in step 2 ($V_{bias}$ = +0.8 V, panel C), where translocation events form a distinct cluster. Linear events are highlighted in blue and the τ vs. event index plot (inset) again shows no apparent correlation. Surprisingly, in step 3 ($V_{bias}$ = -0.8 V, panel D), translocation occurs in different phases. In phase 1, a cluster of DNA-related events emerges at shorter τ (linear and folded events in green and dark grey, respectively), while no correlation is apparent in the τ vs. event index plot (inset, green section). However, as DNA continues to translocate into the nanopipette, τ not only gradually increases, but seems to transition to a new steady-state value (inset, magenta section). Accordingly, in the cluster plot, the event cluster is nominally shifted to longer τ at comparable $\Delta I_{eff}$ (linear and folded events in magenta and light grey), suggesting that new DNA requires more time to translocate. This could be due to a decrease in the local electric field, but given that the pore current does not change significantly, we consider this unlikely. However, in line with our discussion above, namely that the local concentration of DNA in the pipette tip may be enhanced, we hypothesise that the slowing of the translocation process could be due to increased friction inside the nanochannel, i.e. between the incoming DNA strand and those already present. Notably, we consistently make similar observations with other, longer DNA samples in this study, an aspect we will return to below.

Further analysis of the τ distribution for steps 2 (blue) and 3 (green/magenta) are shown in panel E, including Gaussian fits. In step 2, the τ distribution is shifted towards longer times and is broader ($\tau_{max}$ = 2.0 ms; $\sigma/\tau_{max}$ = 0.18), while for step 3, $\tau_{max,1}$ = 1.1 ms ($\sigma_1/\tau_{max,1}$ = 0.16) in phase 1, and $\tau_{max,2}$ = 1.4 ms ($\sigma_2/\tau_{max,2}$ = 0.13) in phase 2. Thus, despite the differences in position and width, the relative standard deviations, $\sigma_1/\tau_{max}$, is comparable in all three cases.

Finally, the number of DNA molecules remaining in the nanopipette during each step is shown in panel F. This number initially increases (steps 0-3), similar to what has been observed for the 4 kbp DNA above, but then seems to be oscillating around a plateau from step 3 onwards. Hence, while DNA is still transported into and out of the nanopipette, the uptake capacity seems to be limited, consistent with the above hypothesis that DNA remains trapped in the nanochannel. Interestingly, we observed the same qualitative behaviour, as well as some intriguing differences, for 10 kbp and 48.5 kbp DNA.

Before exploring the latter aspects, we further analysed other electric characteristics of the sensor before and after the transition from phase 1 to phase 2, specifically the pore conductance $G_{pore}$ and the electric noise from the "DC" and "AC" channels of the setup, respectively (see Methodology). In this context, it is worth reiterating that, due to the design of our amplifier with two output channels, the "DC" channel effectively contains the steady-state (open) pore current, hence $G_{pore} \approx I_{DC}/V_{bias}$, while the "AC" channel records fast modulations of the current, including typical translocation events. Accordingly, in fig. 4 A, for the different DNA lengths used in this study, we show $G_{pore}$ before (left) and after (right) the transition between different phases (top) as well as the standard deviation of the current in the "AC" channel (bottom). This includes transitions from phases 1 to 2 and, for 48.5 kbp DNA, also the transition to a third phase, as discussed below.

Focusing on the transition from phase 1 to 2 initially, it is notable that despite the emergence of a second translocation cluster, cf. fig. 3 D, no significant change in either $G_{pore}$ or AC noise was observed. With the nanopore current largely unchanged, it is thus unlikely that the slowing of the translocation process in phase 2 is due to a change in the local electric field distribution and more likely to do with friction effects, as suggested above. For 48.5 kbp DNA, we further observed a second transition ("48.5 (II)" in panel A), where DNA translocation was no longer observed ("block"). This was accompanied by a significant (23%) drop in $G_{pore}$ and a tripling of the AC noise, indicative of a significant change in the sensing region of the nanopipette.[30] We return to a more detailed analysis below. Notably, however, $G_{pore}$ does not drop to zero, which could mean that the remaining pore current is due to continued ion transport through a DNA-rich region in the tip, leakage current through the thin quartz walls,[31] or indeed a combination of both. ~~The three phases~~

Taken together, however, it seemed that for longer DNA the transition from phase 1 to 2 and potentially to phase 3, where applicable, was reached sooner than for shorter DNA. This led us to investigate in a more systematic manner the relationship between the DNA length, the number of DNA molecules resident in the nanopipette at the 1/2 transition and $G_{pore}$. Hence, panel B shows a plot of the total number of DNA molecules in the nanopipette up to the end of the bias sequence (4 kbp DNA, open circles) and to the transition from phase 1 to 2 (7, 10 and 48.5 kbp DNA (solid circles)). The numeric label associated with each data point is the value of $G_{pore}$ (in nS), as a proxy for the nanopipette size (see Methods and SI, section 6).

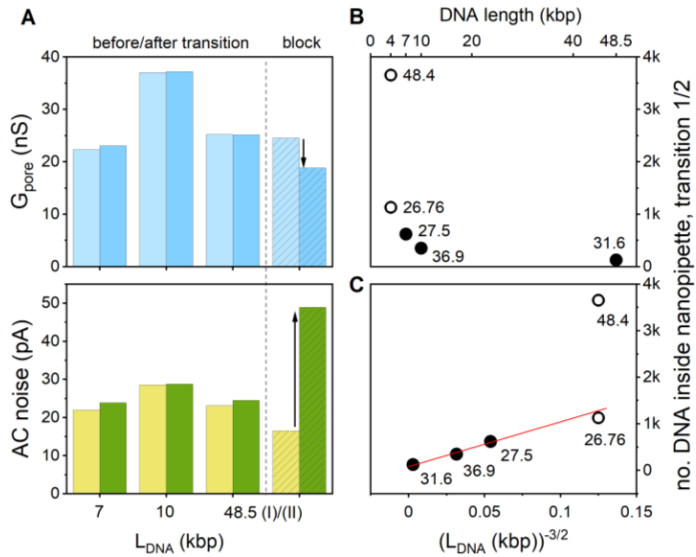

Figure 4. A) $G_{pore}$ (DC channel, top) and electric (AC channel, bottom) before and after the transition from phase 1 to phase 2, left/light blue and right/dark blue, for 7, 10 and 48.5 kbp DNA ("48.5 (I)"). "block" ("48.5 (II)") refers to a second transition observed for 48.5 kbp DNA, see main text. B) Plot of the no. of DNA molecules in the nanopipette at the end of the bias sequence (4 kbp DNA) or until the transition point from phase 1 to phase 2 is reached (7, 10 and 48.5 kbp DNA). The labels indicate $G_{pore}$ of the nanopipette (in nS), as determined by I/V spectroscopy prior to the experiment. C) Plot of the number of DNA molecules inside the nanopipette at the transition from phase 1 to 2 vs. $(L_{DNA})^{-\frac{3}{2}}$. A linear relationship is consistent with space-limiting model of translocation into the nanopipette, as discussed in the main text.

Taking the two results for the 4 kbp DNA sample first, it is apparent that after the same duration, more DNA molecules have translocated into the larger nanopipette (48.4 nS), compared to the smaller one (26.76 nS), as expected in the transport-limited regime.[32] The transition to phase 2 had not occurred in either of the two cases under the experimental conditions used.

For the longer DNA samples in nanopipettes with comparable $G_{pore}$, the number of DNA molecules required to reach the 1/2 transition indeed decreases with length, from 619 for 7 kbp to 349 for 10 kbp and 124 for 48.5 kbp. This observation may be rationalised based on a simple model. To a first approximation, we assume that the nanochannel close to the pipette tip is characterised by an effective, finite volume, $V_{ch}$, that may be filled by DNA with a molecular volume $V_{DNA}$. Treating the DNA as a worm-like chain and $V_{DNA} = \frac{4}{3}\pi R_H^3$, the number of DNA in $V_{ch}$ is then simply $N_{DNA} = V_{ch}/V_{DNA} = \frac{3V_{ch}}{4\pi \cdot R_H^3} = 4.22 \cdot \frac{V_{ch}}{[P \cdot d_{bp}]} \cdot (N_{bp})^{-\frac{3}{2}}$, where $R_H$ is the hydrodynamic radius, P persistence length, $N_{bp}$ the number base pairs, and $d_{bp}$ the average distance between two successive ones, see SI section 7 for further details. This expression predicts a linear correlation between $N_{DNA}$ and $(N_{bp})^{-\frac{3}{2}}$ with slope $4.22 \cdot \frac{V_{ch}}{[P \cdot d_{bp}]}$, which is indeed consistent with the results displayed in panel C. A linear least-square fit yields a slope of $(3.0 \pm 0.4) \cdot 10^8$ bp$^{3/2}$/nm$^3$ (intercept: $80 \pm 40$). Taking P = 50 nm and $d_{bp}$ = 0.34 nm, this provides $V_{ch} \approx 5.1 \cdot 10^9$ nm$^3$ or 5.1 μm$^3$. Approximating the nanochannel tip as a cone with base radius r, height h and opening angle 2α, its volume becomes $V_{cone} = \frac{1}{3}\pi r^2 h = \frac{1}{3}\pi h^3 \tan^2(\alpha)$. Setting $V_{ch} \approx V_{cone}$, $\alpha \approx 5°$, and solving for h yields 8.6 μm (r ≈ 0.75 μm), suggesting that the extension of the nanoconfined region may be significantly smaller than the taper length of the pipette (≈ 3 mm, see SI, table S1).

Extrapolation towards 4 kbp DNA suggest that, for similar $G_{pore}$, the critical limit of ≈ 1300 molecules was not reached under the experimental conditions used here and therefore no 1/2 transition observed. Hence, while the model is relatively simple, it seems to capture some essential features in the experimental data. In future, it could be refined further, e.g. by more accurately describing the DNA polymer in confinement, and complemented with systematic experimental studies, e.g. involving nanopipettes with different shape and geometry in the nanochannel.

However, here we return to a more detailed analysis of the translocation data for 48.5 kbp DNA, fig. 5 ($V_{bias}$ = -0.5 V, $c_{DNA,out}$ = 300 pM). As shown in panel A, DNA translocation initially results in a well-defined cluster of events at $\tau \approx$ 4-11 ms and $\Delta I_e \approx$ 100-200 pA (cf. "phase 1" in panel C). The end of phase 1 is marked by a small increase in $G_{pore}$ of $\approx$ 0.8 nS (3%) and an approximate doubling of the noise in the AC channel from about 10 pA to 22 pA. Subsequently, a second cluster of translocation events began to emerge at $\tau \approx$ 10-20 ms and similar $\Delta I_e$, while $G_{pore}$ and AC noise remained approximately constant (cf. "phase 2" in panel C). The end of this phase is marked by a significant (23%) decrease of $G_{pore}$ (cf. "48.5 (II)" in fig. 4A), an initial spike and then a new steady-state value of the AC channel noise of $\approx$ 50 pA ("phase 3", panel C). During this third phase, no further DNA translocation could be detected over 800 s recording time, as shown in the scatter plot in panel B. Taken together, it appears that DNA initially translocates in an uninhibited fashion (phase 1), then enters a phase where the speed of DNA translocation is reduced, most likely to DNA crowding inside the pipette tip (phase 2) and finally translocation eventually ceases when the nanochannel becomes too densely populated to accommodate more DNA molecules (phase 3). Notably, this effect was reversible in that, when the bias was reversed, DNA was transported out of the nanopipette and the original $G_{pore}$ and AC noise values were recovered (see SI, fig. S10, for data with 10 kbp DNA).

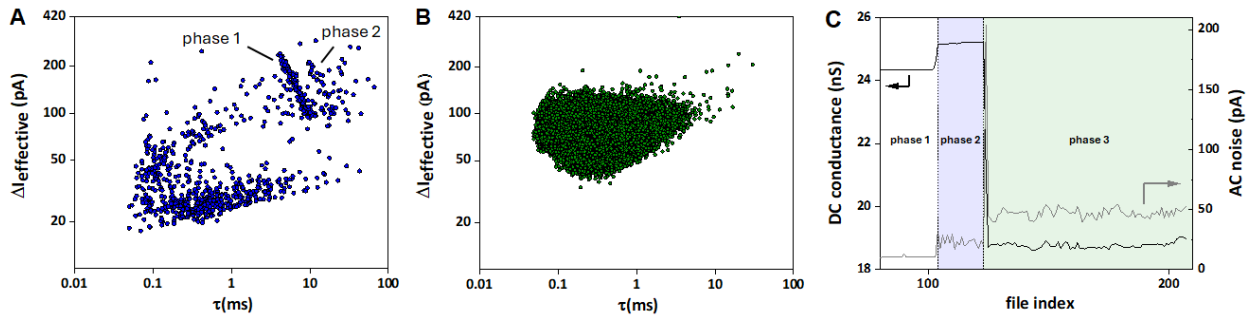

Figure 5, A/B) Scatter plots of $\Delta I_e$ vs. $\tau$ for phases 1 + 2 (A) and phase 3 (B) of a translocation experiment with 48.5 kbp λ-DNA ($V_{bias}$ = -0.5 V, event detection threshold: 3σ). C) $G_{pore}$ and the current noise in the AC channel, as a function of file index (1 file = 10 s run time). Phase 2 is highlighted in blue and corresponds to the emergence of a second cluster of translocation events (panel A). In the "blocking" phase (green), no further translocation events could be detected, as shown in panel B.

When decreasing the DNA concentration outside the nanopipette from $c_{DNA,out}$ = 300 pM to 40 pM ($G_{pore}$ = 44.1 nS), translocation experiments over the same time duration did not lead to the same blocking behaviour, cf. fig. S8 in the SI. In light of the above discussion, this is unsurprising: the lower concentration leads to a significant reduction in the translocation frequency for transport into nanopipette and even if transport within the nanochannel is restricted, the number of resident DNA molecules does not reach the critical value.

The results presented so far provide compelling evidence for the prolonged presence of DNA inside the nanochannel and its effect on the behaviour and operational characteristics of the nanopipette during resistive-pulse sensing experiments. The combination of reversible loading, local storage and unloading of the DNA however raises interesting prospects for new sensor applications. For example, DNA functionalised with appropriate captures probes ("carrier DNA") could initially be kept outside the nanopipette, while their biomolecular targets are largely confined to the inside. The DNA could then be translocated inwards, incubated with the target in the confined region for a suitable amount of time, and finally translocated in the reverse direction to detect where and how many targets are bound to the carrier. The benefits include not only the smaller sample volume on the inside of the pipette and potentially operation in asymmetric electrolyte conditions, but also take advantage of the somewhat increased linear-to-folded ratio for DNA translocation in this direction (which simplifies readout of the capture probe locations), *vide supra*. We demonstrate this approach in a proof-of-concept experiment below. In this instance, we chose gold nanoparticles as targets, as their dispersion in the translocation buffer was found to be sufficiently stable and their relatively large size (compared to some protein targets, for example) rendered them easily detectable.

Specifically, for the carrier we prepared 10 kbp DNA functionalised with a biotin group at one end, cf. SI section 1, and added it to the electrolyte solution (4 M LiCl) on the outside of the pipette. The solution on the inside of the pipette contained streptavidin-functionalised Au particles in the same electrolyte (Nanopartz Inc.; diameter: 40 nm, $c_{particle,in}$ = 500 pM), but initially no DNA. We then translocated ~~the functionalised~~ DNA into the pipette for 1000 seconds ($V_{bias}$ = -0.7 V; $G_{pore}$ = 26.1 nS), switched the applied voltage bias off and left the sample to incubate for ~20 minutes. Subsequently, the bias was reversed for 1000 seconds ($V_{bias}$ = +0.7 V) and the DNA/particle mixture translocated from the inside to the outside of the pipette ("reverse translocation"). Our expectation ~~hypothesis~~ was that ~~during reverse translocation of the DNA/nanoparticle complex,~~ due to the location of the biotin capture probe, translocation events of the DNA/nanoparticle complex would feature nanoparticle-related sub-events ~~would predominantly occur~~ either at the beginning or at the end of a translocation event (depending on in which direction the DNA enters the pore),[33] thereby confirming the successful binding and detection of the target analyte.

This is indeed borne out in the results shown in figure 6. Specifically, panel A shows a scatter plot of the maximum event current $\Delta I_{max}$ vs. $\tau$. To this end, $\Delta I_{max}$ was chosen over $\Delta I_{eff}$, in order to emphasize the difference between events with and without sub-events. Background noise-related events have been removed for clarity.

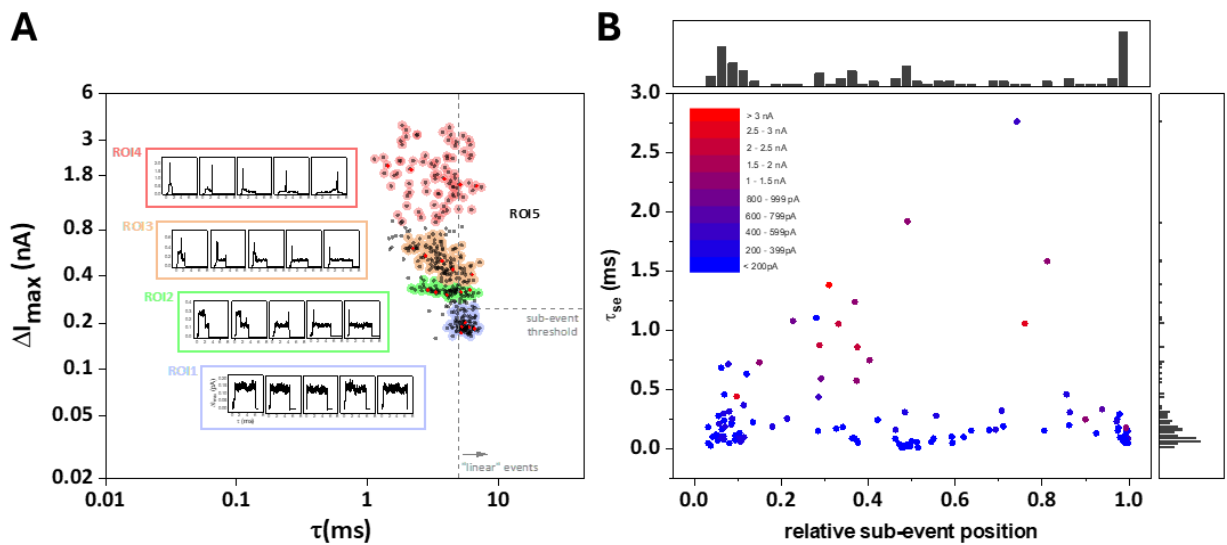

Figure 6: Reverse translocation of biotin-functionalised, 10 kbp DNA and 40 nm streptavidin-modified nanoparticles (4 M LiCl electrolyte, $G_{pore}$ = 26.1 nS; $V_{bias}$ = +0.7 V) A) Scatter plot of $\Delta I_{max}$ vs. $\tau$. Regions-of-Interest (RoI) 1 to 5 are indicated with representative examples (inset), see main text for further discussion. B) For events in RoI5, sub-event duration $\tau_{se}$ vs. relative sub-event position within an event (0: event start, 1: event end). Colour code: sub-event magnitude, from < 200 pA (blue) to > 3 nA (red). Top and right-hand side: histograms of the relative sub-event position and sub-event duration.

Not unpexpectedly, the $\Delta I_{max}$ vs. $\tau$ plot is more complex than those obtained from the translocation of unmodified DNA, cf. figures 2 and 3. Thus, as a first step, we explored the dataset through manuyally selected Regions-of-Interest (RoI), as indicated, with the aim of a high-level classification of event types. Randomly selected example events are shown in the insets with corresponding data points in red (from left to right). Starting with RoI1 (light blue), these events occur at relatively long $\tau$ and low $\Delta I_{max}$, have no significant sub-structure and are predominantly due to the translocation of linear DNA. RoI2 (green) is characterised by a cluster of events with a relatively wide $\tau$- but remarkably narrow $\Delta I_{max}$ distributions, with $\Delta I_{max}$ values ranging from approximately 300-400 pA. Closer inspection of the event shapes indicates that events at longer $\tau$ tend to be linear DNA translocation events with short, spike-like sub-events. Those at shorter $\tau$ display the characteristic features of folded DNA translocation, where the beginning of an event is characterised by an excursion to approximately twice the $\Delta I_{max}$ value for linear DNA translocation with a significant duration, $\tau_{se}$ (see the two example events on the left-hand side of RoI2). Event shapes in RoI3 (orange) again provide evidence for linear and folded DNA translocation (from longer to shorter $\tau$ values, right to left), but the event magnitudes are typically defined by short, intensive sub-events ($\Delta I_{max} < 0.8$ nA). Finally, RoI4 (light red) encompasses events in a more diffuse cluster that are of similar appearance to those in RoI3, albeit with even larger sub-events ($\Delta I_{max} > 1$ nA). This event magnitude is indeed comparable to the current change observed when the nanopipettes are seemingly blocked with DNA, cf. fig. 5, and we therefore speculate that events in RoI4 may involve pipette blockage with sufficiently large nanoparticles.

However, returning to the original hypothesis, namely whether we can confirm nanoparticle binding to the functionalised (biotinylated) part of the DNA, we chose to define a further RoI, RoI5. To this end, we determined the mean $\tau_{m,1}$ and standard deviation $\sigma_{m,1}$ from a Gaussian fit of the translocation time distribution for the events in RoI1 (linear events only; $\tau_{m,1} = 0.0057$ ms, $\sigma_{m,1}$

= 0.0007 ms) and then considered all events with $\tau > (\tau_{m,1} - \sigma_{m,1})$ and $\Delta I_{max} > 250$ pA as nominally linear DNA translocation events with sub-events. In this way, we identified 63 events from RoI1-4, which were then subjected to a sub-event search. This identified 115 sub-events in total, namely 34 events with one sub-event, 17 with two, 9 with three and two with four. One event with one significant sub-event feature as well as two others were misclassified; a broad selection of individual events is shown in the SI, section 10, for reference. Fig. 6B shows the results of this analysis as a plot of $\tau_{se}$ vs. the relative sub-event position within an event (0: event start, 1: event end), along with the respective histograms at the top and the right-hand side. The colour code represents the sub-event intensity, from < 200 pA (blue) to > 3 nA (red). From this analysis, it becomes apparent that the majority of sub-events occurs close to the start or the end of an event (66 out of 115 within the first or last 15% of the event), that the dominant $\tau_{se}$ is relatively short (mode = 60 µs, see $\tau_{se}$ histogram) and their magnitude low (< 400 pA). Hence, this statistical distribution of sub-events indeed provides evidence that the nanoparticles are bound to the DNA and detected successfully in this experimental configuration. Interestingly, there also appears to be a somewhat increased probability of low-magnitude, short sub-events close to the central region of translocation events (relative position ≈ 0.5), even though there is no obvious streptavidin binding site in this part of the functionalised DNA. While statistically the simultaneous translocation of DNA and nanoparticles ("co-translocation") is not unexpected, it is not immediately obvious why nanoparticles should more likely co-translocate in any particular region of an event and should be more evenly distributed. Possible reasons could be related to specific DNA/nanoparticle interactions or the dynamics of the translocation process, but this aspect requires further study.

CONCLUSIONS

In conclusion, we have shown through a range of experiments how confinement in the nanopipette can influence the translocation of kbp DNA into and out of the nanopipette tip. Slow transport from the tip region into the bulk of the nanopipette, likely due to a combination of spatial constraints, friction and weakening electric field, appear to be important in this context. Specifically, after loading the nanopipettes with DNA, the translocation frequency for reverse translocation (out of the pipette) was much higher than expected, based on estimates for homogenous mixing, and compared to inward translocation (i.e., the local concentration in the tip was enhanced). In line with previous studies, we found inward translocation to be faster than outward translocation, while the variance of the translocation time distribution scaled accordingly. We also observed a somewhat larger fraction of linear DNA when translocating out of the pipette, for all DNA lengths studied here, which again may be related to confinement inside the nanopipette tip.

To this end, confinement effects were particularly prominent for longer DNA, where we observed that the translocation process undergoes different stages, depending on the number of DNA molecules resident in the pipette. While initially all DNA samples displayed unperturbed translocation into the pipette ("open" state), with well-defined translocation time distributions for the translocation of linear and folded DNA, the system was found to switch to a new translocation state once a critical number of DNA inside the pipette – and by implication, inside the nanochannel – was reached ("crowded" state). This limiting value was smaller for longer DNA for comparable nanopipette sizes and we have presented evidence that this is consistent with a "finite volume" model of the tip region. While in our study all nanopipettes were fabricated with similar pulling parameters and displayed comparable $G_{pore}$ values (with the exception of one dataset for 4 kbp

DNA, *vide supra*), we envisage that $V_{ch}$ could be systematically varied by changing the pore diameter, the internal dimensions and shape of the nanochannel ~~(e.g., via the taper length of the pipette)~~ in future studies, to further test our hypothesis and potentially refine the model.

For the 10 kbp and 48.5 kbp DNA, we moreover observed that after continued operation, DNA translocation eventually ceased and that the nanopipette reached a "blocked" state. This was accompanied by a marked drop of the current in the DC channel (albeit not to zero) and a significant increase of the electric noise in the AC channel. Notably, the transitions between the open, crowded and blocked states were found to be reversible, which inspired a proof-of-concept experiment exploiting the crowding effect for sample processing and incubation. For this purpose, we filled the nanopipette with a solution containing streptavidin-modified, 40 nm gold nanoparticles, translocated biotin-functionalised DNA into the nanopipette and, after 20-minute incubation, reverse translocated the DNA out of the pipette. Analysis of the translocation data allowed for the identification of various types of events associated with bare DNA, DNA-nanoparticle complexes as well as seemingly simultaneous translocation of DNA and nanoparticles ("co-translocation"). Importantly, we were able to demonstrate through detailed analysis of the event sub-structure that the nanoparticles can be bound specifically to the DNA and that target capture has indeed been successful. This integration of sample incubation and detection into the nanopipette therefore suggests a new bioanalytical paradigm, which could be extended to other targets such as proteins or RNA and different electrolytic media. The latter could include asymmetric configurations, i.e. with different solutions on the in- and outside of the nanopipette, adding substantial flexibility and generality to this analytical approach.

METHODS

Ag/AgCl electrodes were freshly prepared using anodization as previously described.[20] First, 10 cm of silver wire (0.25 mm diameter, 99.99% purity, Goodfellow) was cut and immersed in 38% v/v nitric acid (Sigma) for 10 seconds, then washed with Milli-Q water (18 M$\Omega$, Merck Millipore) to remove surface impurities. The cleaned wires were soldered to gold contact pins and submerged in the 4 M LiCl 1xTE solution. Anodization was performed in an electrochemical cell using a gold wire (99.99% purity, Goodfellow) as counter electrode and applying a current of 1 mA until the electrode surface turned black/purple.

Nanopipettes were fabricated using laser mechanical pulling of plasma-cleaned filamented quartz capillaries (outer diameter: 1 mm, inner diameter: 0.5 mm, length: 7.5 cm, Sutter Instruments, Novato, USA) with a P2000 pipette puller. The pulling parameters were optimized using two lines: line 1 (heat: 700, filament: 5, velocity: 35, delay: 150, pull: 75) and line 2 (heat: 700, filament: 0, velocity: 15, delay: 128, pull: 200). To mitigate thermally induced inconsistencies in pore size, the room temperature was maintained at approximately 20°C. The taper length of the nanopipettes was subsequently determined using optical microscopy (Nikon, T-105) and ranged from 3.0 to 3.2 mm in the present study, cf. SI section 6.

The pore diameter $d_p$ was determined from the slope of the I/V curve measured in 4 M LiCl electrolyte solution using a custom-designed cell, based on equation 1:[23]

$$d_p = \frac{4GL + \frac{\pi}{2}G_{pore}D_i}{\pi D_i g - \frac{\pi}{2}G_{pore}} \quad (1)$$

where $G_{pore}$ is the conductance of the pore, l the taper length of the nanopipette (by optical microscopy), $D_i$ is the inner diameter of the capillary (0.5 mm, as per manufacturer) and g is the conductivity of the electrolyte solution (173 mS cm$^{-1}$).[20] Values obtained for $d_p$ were between 12 and 23 nm in this study, cf. SI section 6.

To prevent or reduce the formation of air bubbles, the electrolyte solution was slowly filled back and forth into the nanopipette using a Microfil™ needle. Only nanopipettes displaying an Ohmic current-voltage (I/V) relationship were used, while those that did not meet this criterion were discarded. Ohmic behavior was defined using the ion current rectification ratio (r), calculated as r =|I-/I+|,[34] at ±0.5 V. Nanopipettes with r values between 0.9 and 1.2 were considered acceptable.[34]

For DNA translocation experiments. 4 kbp, 7 kbp, 10 kbp (Thermo Fisher, NoLimits DNA), 48.5 kbp (NEB) and biotinylated 10 kbp DNA PCR product (detailed synthesis in the SI, section 1) were injected separately to the bulk solution in a custom designed liquid cell containing 0.1-3 mL of 4 M LiCl (bulk DNA concentrations 40- 600 pM, as indicated). The liquid cell was housed in a double Faraday cage to reduce electric interference. A negative bias value means that the electrode outside the nanopipette is biased negatively, thereby resulting in an electrophoretic driving force for (negatively charged) DNA to translocate into the pipette.

Experiments were conducted in a semi-automated fashion using in-house MATLAB code with a sequence of applied biases, where for each bias 100 data files of 10 s each were collected before the next bias value was applied. Under the electrolyte conditions used, typical DNA translocation leads to a decrease of pore current and events with negative polarity. However, for convenience, we use the absolute event magnitudes throughout this work.

Data recording was performed at a sampling rate of 1 MHz using custom-built low-noise, high-bandwidth amplifier connected to digital oscilloscope for analog-to-digital conversion (Picoscope 4262, Pico Technology), as reported previously.[5, 6] Briefly, in this design the input current is split into two output channels, namely the "DC" and "AC" channels. The former contains slow modulations of the input current (cutoff frequency ~ 7 Hz), including the steady-state pore current which may be used to estimate $G_{pore}$. The AC channel contains fast modulations of the input current, for example (short-lived) translocation events, and is normally zero mean, facilitating baseline correction. For the results presented here, the AC channel output was filtered using a eight-pole low-pass (analog) Bessel filter (cutoff: 100 kHz, Krohn-Hite Corporation) and, in some cases, also digitally filtered as indicated.

Event detection

Data analysis involved zero-order baseline background correction of the AC channel output to remove minor current offsets (<10 pA), when necessary. Using custom-built MATLAB code, events were detected with a 5σ threshold, unless specified otherwise, where σ was the standard deviation of the AC channel noise. For each detected event, relevant segments of the current-time trace were extracted from and up to the adjacent zero crossings and relevant event characteristics determined, such as the event current magnitude, $\Delta I_{max}$ and the effective event current, $\Delta I_e$ (from the event charge and its duration). Additional characteristics included the event duration based on 1σ threshold crossings (τ), which we found to better capture the characteristics of events with complex shapes, or the event duration at full-width half-maximum.[5, 20] For sub-event detection, a separate threshold search was performed on each event with a threshold value of 200 pA, relative to the median baseline of the event (calculated between 0.1 and 0.9 of the relative event duration), and the number, position and other sub-event characteristics extracted.

The separation of linear and folded event populations was conducted by first extracting the overall DNA event cluster in $\Delta I_{eff}$ or $\Delta I_{max}$ vs. $\tau$ scatter plot using DBSCAN. For these events, five features, namely the event duration ($\tau$), the effective current ($\Delta I_e$), the maximum event current ($\Delta I_{max}$), the AC channel noise, and the event charge (q), were first standardised (z-score) and then subjected to Principal Component Analysis (PCA). The first two principal components were retained and subsequently separated using k-means clustering with two populations. The cluster with the longer $\tau_m$, as determined from a Gaussian fit of the respective $\tau$ distribution, was associated with linear DNA translocation events and colour-coded in the corresponding $\Delta I_e$ or $\Delta I_{max}$ vs. $\tau$ plots, as indicated. Further information can be found in section 4 of the SI.

ASSOCIATED CONTENT

**Supporting Information**.

The following files are available free of charge.

Supporting Information with further details on DNA preparation protocols, gel electrophoresis data, additional translocation data not included in the main manuscript and representative examples of translocation events for the DNA/nanoparticle experiment shown in fig. 6 (file type: PDF)

AUTHOR INFORMATION

**Corresponding Author**

* Tim Albrecht, University of Birmingham, School of Chemistry, Edgbaston Campus, Birmingham B15 2TT, United Kingdom, t.albrecht@bham.ac.uk

**Author Contributions**

The manuscript was written through contributions from all authors. All authors have given approval to the final version of the manuscript.

**Funding Sources**

Part of this work was funded by the Leverhulme Trust (RPG-2022-165). LM wishes to acknowledge support for a joint PhD studentship from the University of Birmingham, UK, and the Federal Institute for Materials Research and Testing, Berlin/Germany. RAAW acknowledges financial support from the Jordan University of Science and Technology.

**Notes**

Table of Contents figure

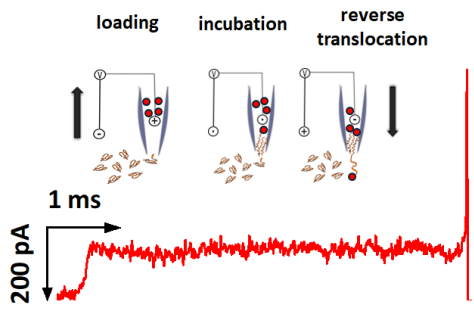

# Crowding Effects during DNA Translocation in Nanopipettes

*Rand A. Al-Waqfi[§,£], Cengiz Khan[§], Oliver J. Irving[§], Lauren Matthews[§,&], Tim Albrecht[§]*

[§] University of Birmingham, School of Chemistry, Edgbaston Campus, Birmingham B15 2TT, United Kingdom

[£] Department of Medicinal Chemistry and Pharmacognosy, Faculty of Pharmacy, Jordan University of Science and Technology, P.O. Box 3030, Irbid 22110, Jordan

[&] Federal Institute for Materials Research and Testing, Department 6, Unter den Eichen 87, 12205 Berlin, Germany

# Supporting Information

1. **Synthesis of 10 kbp biotin-functionalised DNA:**

Synthesis of 5'biotin-functionalised DNA was conducted using polymerase chain reaction (PCR). The forward (5'-TCATCAGGGCGAGATGCTCAATG) and reverse primer (5'-[biotin]-AAGGCGTTTCCGTTC TTCTTCGT) were provided by Integrated DNA Technologies (IDT) and reconstituted to a concentration of 10 µM in nuclease-free water. Briefly, the reaction was carried out using 25 µL of Q5 High-Fidelity 2X Master Mix (NEB), 2.5 µL of each primer, 1 µL of λ DNA as template DNA (1 ng/µL, NEB) and 19 µL NFW (VWR international). The reaction components were gently mixed by pipetting and briefly centrifuged to ensure homogeneity. LoBind© tube was placed inside PrimeG© thermocycler (Cole Palmer) and thermal cycling conductions were applied as follows: an initial denaturation at 98°C for 30 seconds, followed by 30 cycles of 98°C for 10 seconds, 70°C for 30 seconds, and 72°C for 5 minutes, concluding with a final extension at 72°C for 2 minutes. PCR product was purified using Monarch PCR and DNA Clean-Up kit (NEB) and eluted in nuclease free water. The integrity and expected size of the biotinylated DNA fragment were verified by 1% TAE agarose gel electrophoresis and quantified using a UV-Vis spectrophotometer.

## 2. Gel Electrophoresis results for DNA standards and 10 kbp PCR product:

4 kbp, 7 kbp and 10 kbp DNA samples (NoLimits DNA fragments, Thermo Fisher) as well as 48.5 kbp DNA (NEB) were used. The purity of DNA standards and the success of the above-mentioned PCR reaction were confirmed by gel electrophoresis. In figure S1A, lanes 2-4 showing bands corresponding to 4 kbp, 7 kbp and 10 kbp DNA, respectively.

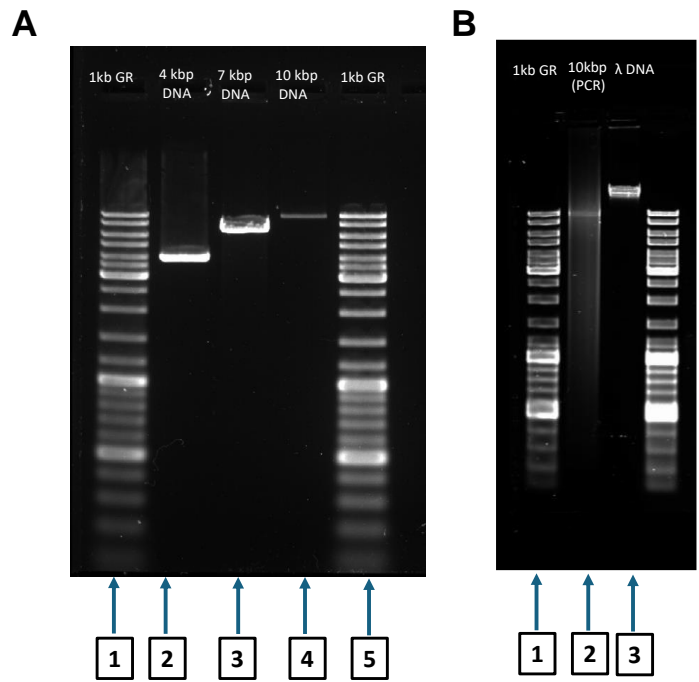

**Figure S1:** Gel electrophoresis results for: A) 4, 7 and 10 kbp DNA samples (lanes 2-4). B) biotin-functionalised 10 kbp PCR product (lane 2) and 48.5 kbp DNA fragment (lane 3). The gels were run for 49 min at 80 V (1% (w/v) agarose gel prepared in 1× TAE buffer).

## 3. Examples of current-time traces (AC channel) with typical translocation events

The traces below are example outputs for 4 kbp DNA from the AC channel (sampling time: 1 µs, low-pass filtered at 100 kHz), see Methods section in the main text. The transient on the left of each measurement results from switching $V_{bias}$ to the specified value. Data in panel A were recorded at $V_{bias}$ = +0.8 V while no DNA was inside the nanopipette and therefore no events observed. The measurement in panel B was recorded at $V_{bias}$ = -0.6 V, resulting in DNA translocation into the nanopipette. Insets: example events, likely showing the translocation of folded and linear DNA, respectively. Finally, data shown in panel C were measured at $V_{bias}$ = +0.6 V. Translocation events are now in the upwards direction, but note that the total current is composed of both AC and DC channels. Thus, under the high-ionic strength conditions used (4 M LiCl), DNA translocation events are decreases in the overall current (blockages).

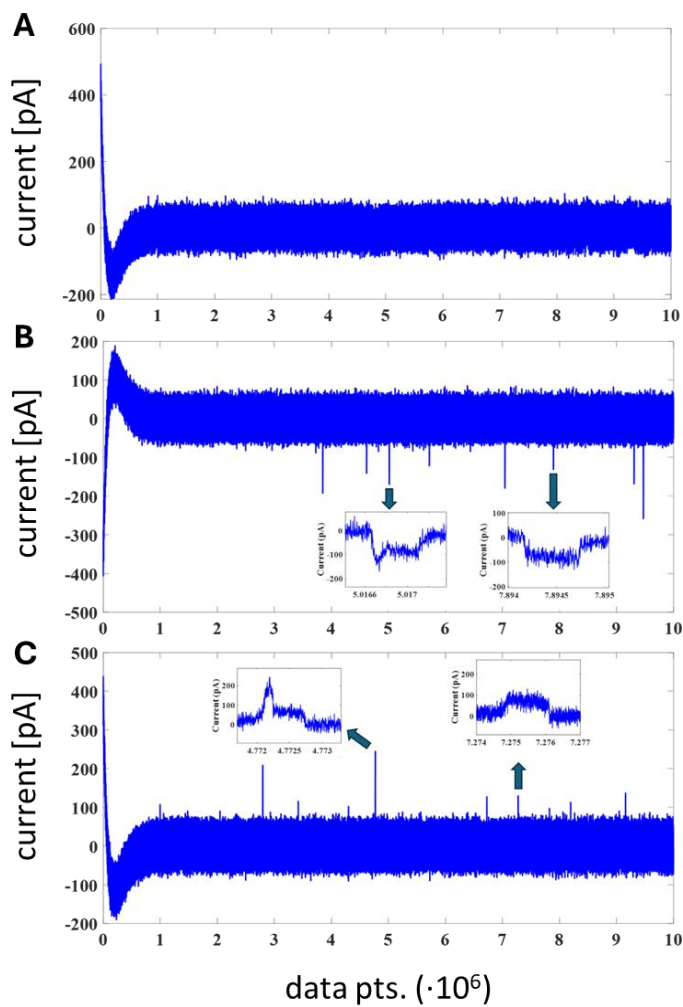

**Figure S2:** Raw data output for 4 kbp DNA (AC channel), current vs. data points. A) $V_{bias}$ = +0.8 V, no DNA translocation occurs as the initial DNA concentration inside the pipette is zero. Panels B/C): $V_{bias}$ = -0.6 V and +0.6 V. DNA events are observed in both cases.

## 4. Separation of linear and folded DNA translocation events – an example

Following the description of the separation of linear and folded DNA translocation events in the main manuscript (cf. Methods), we further illustrate this process here. To this end, fig. S2 (left) shows the $\Delta I_{eff}$ vs. $\tau$ scatter plot from a translocation experiment with 4 kbp DNA ($V_{bias}$ = -0.8 V, 4 M LiCl). A distinct cluster of DNA translocation events is visible and indicated by the ellipse. This subset of events is first extracted using DBSCAN (or another suitable clustering method). After z-score standardisation, PCA analysis is performed based on an expanded set of five features (see main manuscript), the first two principal components retained and subsequently clustered into two sub-populations using k-means clustering. The sub-population formally associated with linear events is colour-coded in red. Two example events for linear and folded DNA translocation are shown on the right.

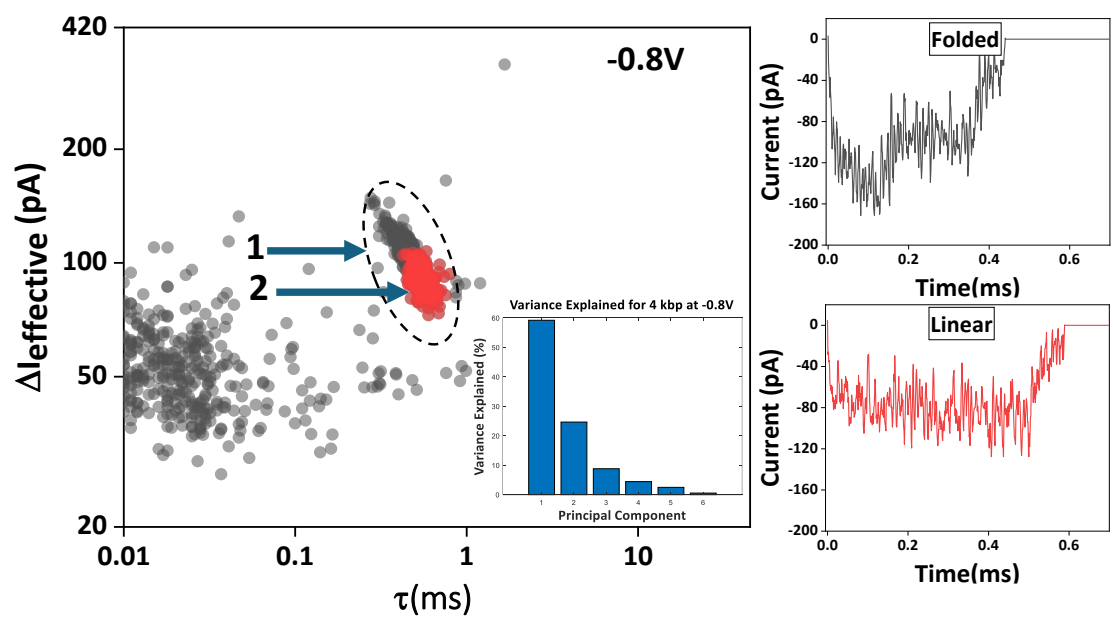

**Figure S3.** Left: scatter plot of $\Delta I_{eff}$ vs. $\tau$ for 4 kbp DNA ($c_{DNA,out}$ = 300 pM, -0.8 V). The overall DNA-related cluster is enclosed within an ellipse and shown in black. Linear translocation events, isolated in step 2 using PCA and k-means clustering, are highlighted in red. The Scree plot (bottom inset) illustrates the variance captured by each principal component, with the first two components explaining ~85% of the total variance. Right: two representative current-time events, namely folded (top, black) and linear (bottom, red).

## 5. Translocation data for other DNA lengths, ΔI$_{eff}$ vs. τ scatter plots (examples)

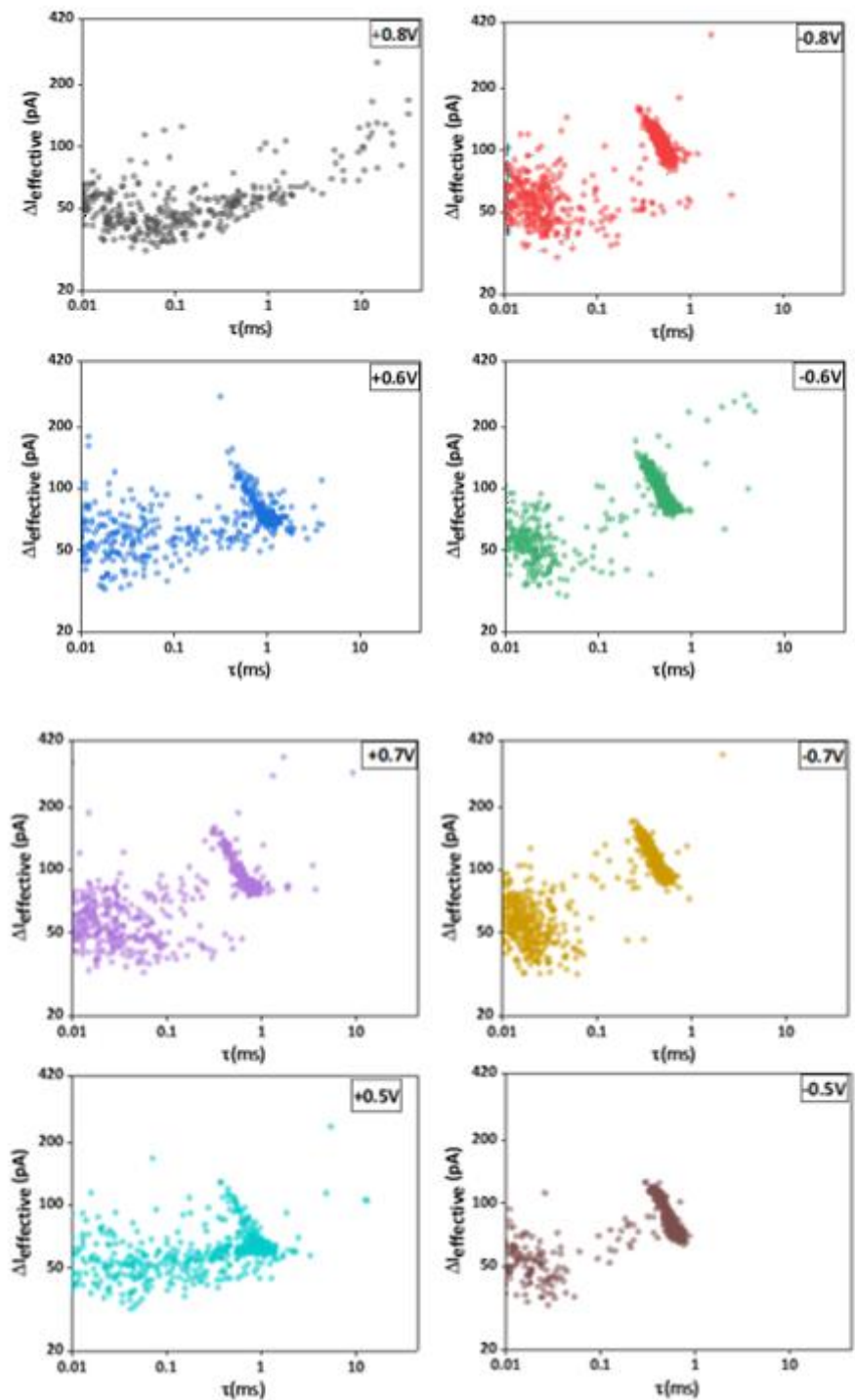

**Figure S4:** Scatter plots of ΔI$_{eff}$ vs. τ for 4 kbp DNA ($c_{DNA,out}$ = 300 pM), bias applied in sequence from left to right. No DNA cluster at +0.8 V, since $c_{DNA,in}$ = 0 at the beginning of the experiment. The applied bias sequence was from left to right, top to bottom.

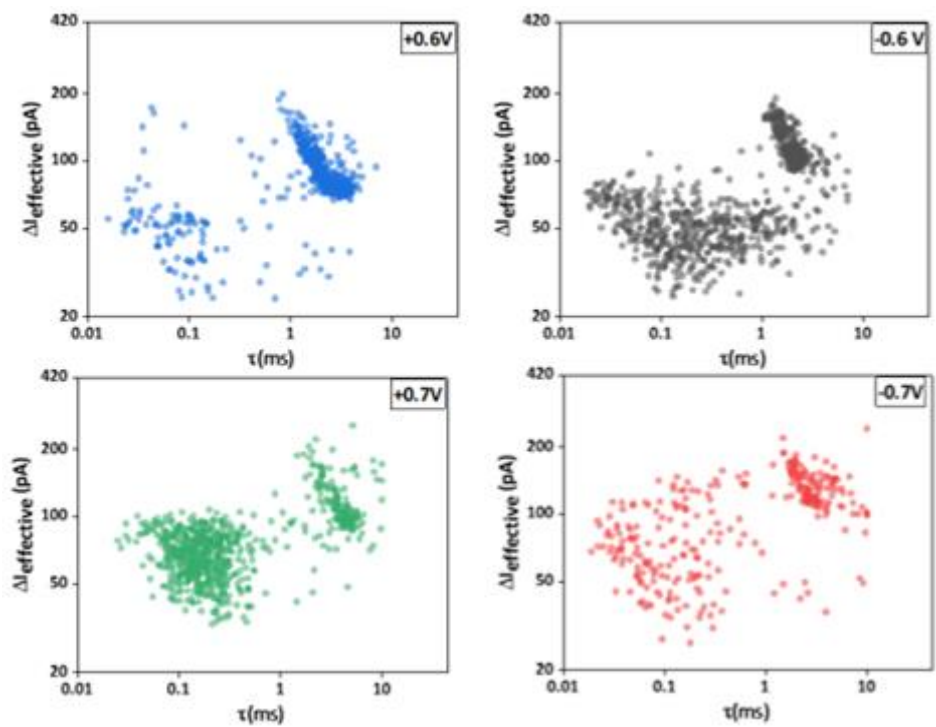

**Figure S5:** Scatter plots of $\Delta I_{eff}$ vs. $\tau$ for 7 kbp ($c_{DNA,out}$ = 600 pM). The applied bias sequence was from left to right, top to bottom.

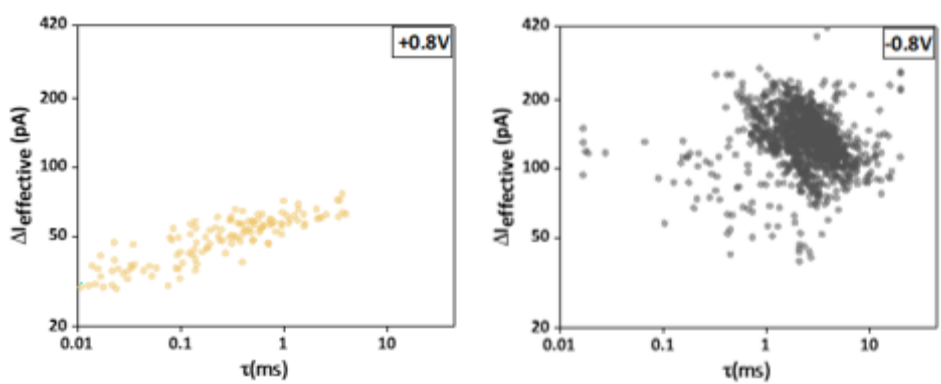

**Figure S6:** Scatter plots of of $\Delta I_{eff}$ vs. $\tau$ for 10 kbp DNA ($c_{DNA,out}$ = 600 pM).

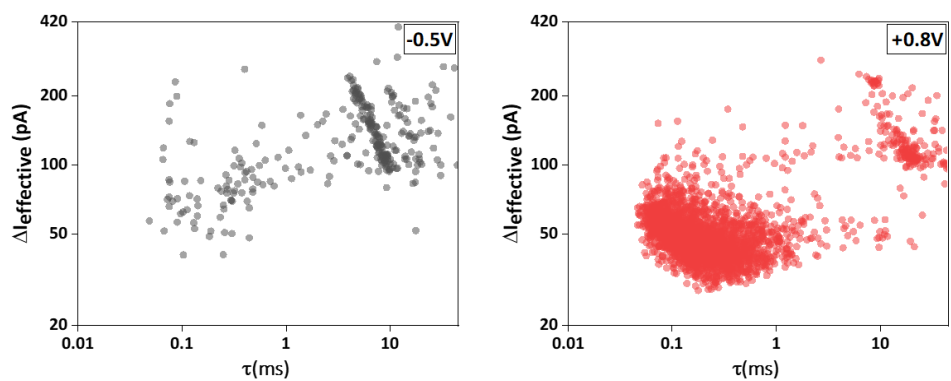

**Figure S7:** Scatter plots of $\Delta I_{eff}$ vs. $\tau$ for 48.5 kbp λ-DNA ($c_{DNA,out}$ = 300 pM). Data were filtered at 10kHz.

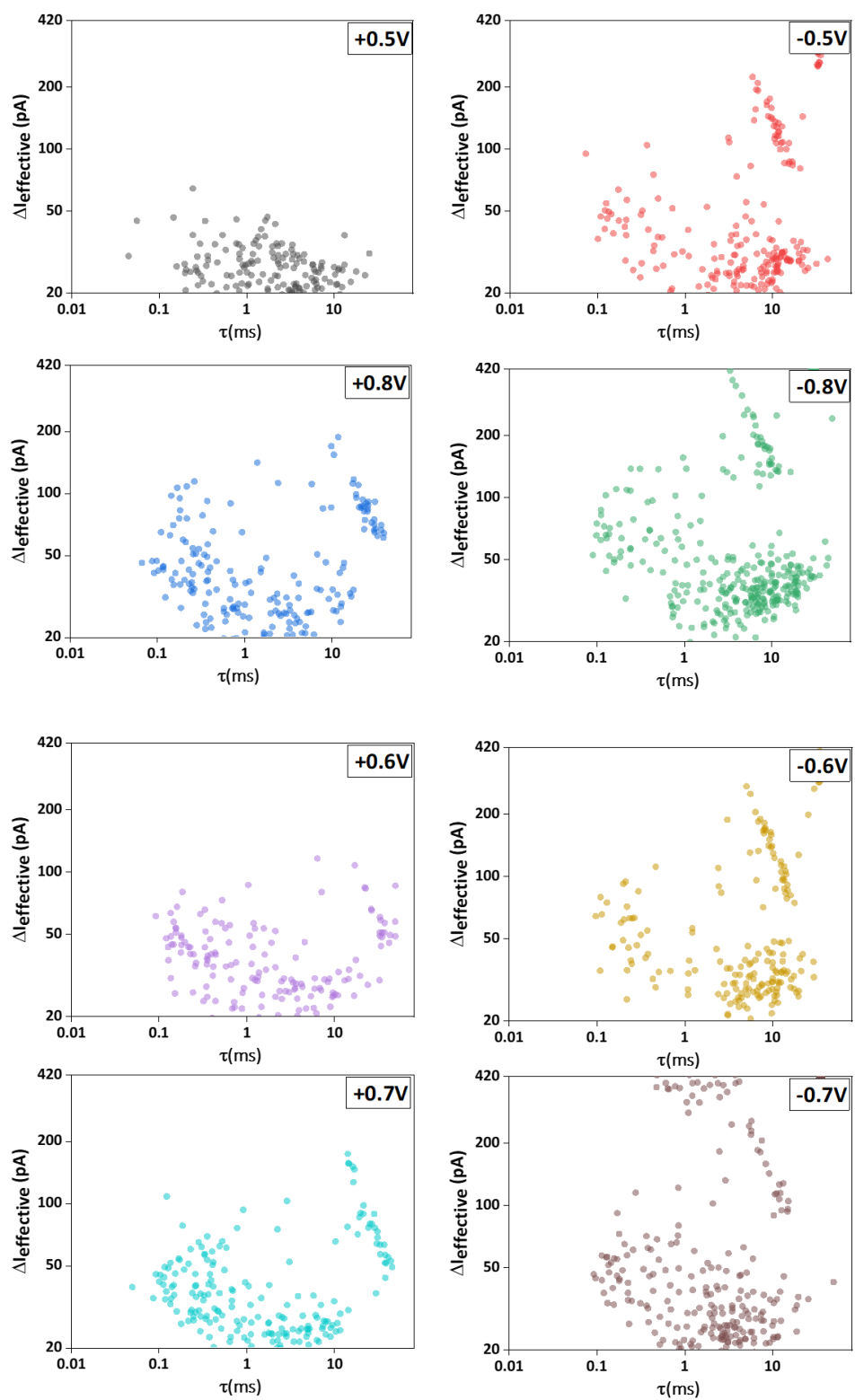

**Figure S8:** Scatter plots of $\Delta I_{eff}$ vs. $\tau$ for $\lambda$ DNA ($c_{DNA,out}$ = 40 pM). The applied bias sequence was from left to right, top to bottom.

## 6. Nanopipette characterisation: physical dimensions and conductance

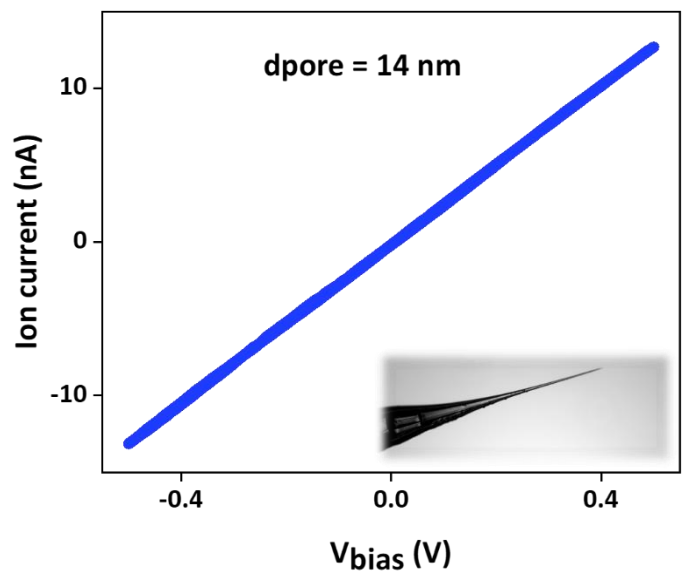

**Figure S9:** I/$V_{bias}$ curve for a typical quartz nanopipette in 4 M LiCl with 10 mM TE buffer. The conductance is G = 27.5 nS, as determined from the average of the slopes from the forward and reverse scan between ±0.5 V. Bottom inset: optical micrograph showing the taper of the same pipette, with a length of approximately 3.17 mm

Table S1: Nanopipette characterisation and DNA concentrations

|                | 4 kbp     | 4 kbp    | 7 kbp    | 10 kbp   | 48.5 kbp  | 48.5 kbp |
|----------------|-----------|----------|----------|----------|-----------|----------|
| **Est. Pore size** | 14 nm     | 20 nm    | 14 nm    | 18 nm    | 12 nm     | 23 nm    |
| **Taper length**   | 3.12 mm   | 3.05 mm  | 3.17 mm  | 3.2 mm   | 3.132 mm  | 3.1 mm   |
| **Conductance**    | 26.76 nS  | 48.4 nS  | 27.5 nS  | 36.9 nS  | 31.6 nS   | 44.1 nS  |
| **DNA conc.**      | 300 pM    | 300 pM   | 600 pM   | 600 pM   | 300 pM    | 40 pM    |

## 7. Estimate of $V_{ch}$ based on a worm-like chain model for DNA

In order to estimate the volume of the nanoconfined region in the nanopipette tip, $V_{ch}$, we assume that $N_{DNA} \cdot V_{DNA} = V_{ch}$ at the transition from phase 1 to 2 during translocation experiments (i.e., that the space is completely filled, albeit not closely compacted). For the description of the DNA, we adopted the worm-like chain model which yields for the mean-square radius of gyration, $R_g$:[1]

$$\langle R_g^2 \rangle = \frac{b^2 N_{Kuhn}}{6}$$

b is the Kuhn length and $N_{Kuhn}$ the number of Kuhn monomers in the chain. $N_{Kuhn}$ is equal to $\frac{N_{bp} \cdot d_{bp}}{b}$ where $N_{bp}$ is the number of base pairs in the strand and $d_{bp}$ the average distance between two adjacent ones. Noting that b is equal to two times the persistence length P, we have:

$$\langle R_g^2 \rangle = \frac{2P \cdot N_{bp} \cdot d_{bp}}{6}$$

Finally, since $\frac{R_H}{R_G} = \frac{3}{8}\sqrt{\pi} \approx 0.665$, we can write an expression for the spherical volume of a DNA strand in terms of the hydrodynamic radius $R_H$ as $V_{DNA} = \frac{4}{3}\pi \cdot R_H^3$.

Since $N_{DNA} = V_{ch}/V_{DNA}$, see main text, we obtain:

$$N_{DNA} = \frac{V_{ch}}{V_{DNA}} = \frac{3 V_{ch}}{4\pi \left(\frac{3}{8}\sqrt{\pi} \cdot R_G\right)^3} \approx 4.22 \cdot \frac{V_{ch}}{(P \cdot d_{bp})} \cdot (N_{bp})^{-\frac{3}{2}}$$

Hence, a plot of $N_{DNA}$ vs. $N_{bp}^{-\frac{3}{2}}$ should show a linear correlation with slope $4.22 \cdot \frac{V_{ch}}{(P \cdot d_{bp})}$, cf. fig. 4C, which based on known values of P and $d_{bp}$ allows for the estimation of $V_{ch}$.

## 8. Additional translocation statistics for the DNA samples studied in this work

Table S2:

| DNA length (kbp) | $V_{bias}$ (V) | ratio, linear vs. folded | $\tau_m$ (linear events) (s) | $G_{pore}$ (nS) | no. of events |
|---|---|---|---|---|---|
| 4 | -0.6 | 0.596 | 0.000538 | 48.4 | 1028 |
| 4 | -0.5 | 0.604 | 0.000596 | 48.4 | 1450 |
| 4 | -0.7 | 0.569 | 0.000480 | 48.4 | 1778 |
| 4 | -0.8 | 0.562 | 0.000442 | 48.4 | 1077 |
| 4 | -0.4 | 0.509 | 0.000571 | 48.4 | 432 |
| 4 | 0.6 | 0.6348 | 0.000816 | 48.4 | 419 |
| 4 | 0.5 | 0.549 | 0.000686 | 48.4 | 271 |
| 4 | 0.7 | 0.6729 | 0.000837 | 48.4 | 691 |
| 4 | -0.5 | 0.629 | 0.000615 | 26.76 | 383 |
| 4 | -0.8 | 0.55 | 0.000551 | 26.76 | 673 |
| 4 | -0.6 | 0.59055 | 0.000568 | 26.76 | 464 |
| 4 | -0.7 | 0.59666 | 0.000486 | 26.76 | 534 |
| 4 | -0.4 | 0.643 | 0.000723 | 26.76 | 253 |
| 4 | 0.6 | 0.7703 | 0.001060 | 26.76 | 283 |
| 4 | 0.7 | 0.834 | 0.000791 | 26.76 | 362 |
| 4 | 0.5 | 0.8 | 0.001070 | 26.76 | 280 |
| 7 | -0.5 | 0.576 | 0.001600 | 27.5 | 520 |
| 7 | -0.8 | 0.538 | 0.001500 | 27.5 | 1069 |
| 7 | -0.6 | 0.556 | 0.002300 | 27.5 | 254 |
| 7 | -0.7 | 0.42 | 0.004000 | 27.5 | 193 |
| 7 | 0.5 | 0.50993 | 0.006500 | 27.5 | 183 |
| 7 | 0.8 | 0.6319 | 0.002000 | 27.5 | 398 |
| 7 | 0.6 | 0.61966 | 0.003000 | 27.5 | 573 |
| 7 | 0.7 | 0.4918 | 0.004600 | 27.5 | 151 |
| 10 | -0.8 | 0.4972 | 0.004040 | 36.9 | 1215 |
| 10 | -0.6 | 0.572 | 0.001640 | 36.9 | 131 |
| 10 | -0.7 | 0.462 | 0.002970 | 36.9 | 1051 |
| 10 | 0.7 | 0.552 | 0.003960 | 36.9 | 1035 |
| 48.5 | -0.5 | 0.483 | 0.008590 | 31.6 | 172 |
| 48.5 | 0.8 | 0.5703 | 0.021520 | 31.6 | 154 |

## 9. Loading, blocking and unloading of nanopipettes: 10 kbp DNA

Translocation of 10 kbp DNA occurred in different phases, from uninhibited translocation (phase 1) to delayed translocation (phase 2), fig. S9 panels A + B, and ultimately blockage (phase 3). Compared to 48.5 kbp DNA, the latter required a larger number of DNA translocation events, but as in the case of the longer DNA, $G_{pore}$ drops and AC channel noise increase significantly. Note that this effect is reversible under the experimental conditions used, as demonstrated by the bias reversal (and unloading) to $V_{bias} = +0.7$ V, panel C.

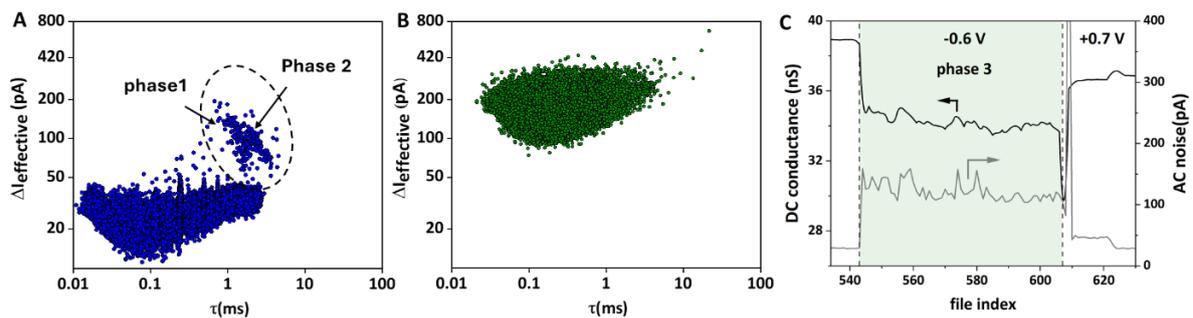

**Figure S10:** A/B) Scatter plots of $\Delta I_{eff}$ vs. $\tau$ for 10 kbp DNA, for the phases 1 + 2 (A) and phase 3 (B) ($V_{bias} = -0.6$ V, event detection threshold of $1.5\sigma$). C) $G_{pore}$ and standard deviation of the current noise in the AC channel, as a function of file index (1 file = 10 s run time). Phase 3 is highlighted in green. The final set at $V_{bias} = +0.7$ V shows the reversal of the blocking, with a recovery of $G_{pore}$ and current noise after DNA is transported out of the nanopipette.

## 10. DNA/nanoparticle translocation: example events with sub-event structure

While the event threshold was at 5σ, where σ is the standard deviation of the current noise in the AC channel (see Methods), events below are plotted from the last zero crossing before event (left-hand side) to first zero recrossing after the event (sharp drop, right-hand side). The threshold for sub-event detection was 200 pA, relative to the median of the event baseline between 0.1 and 0.9 relative event time.

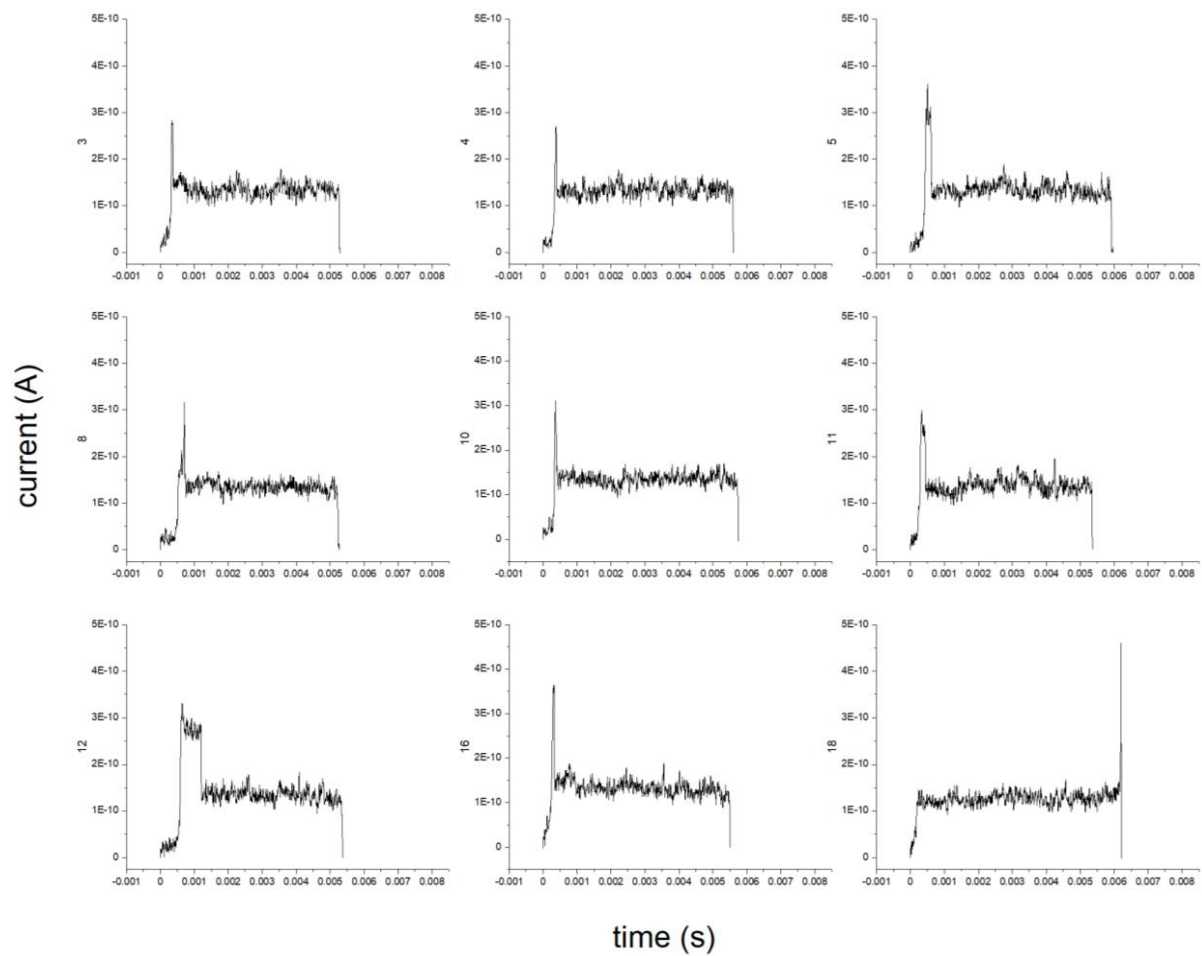

**Figure S11**: Current-time traces for the first 9 events formally identified with **one** sub-event. The current range is from -25 pA to 500 pA for all events.

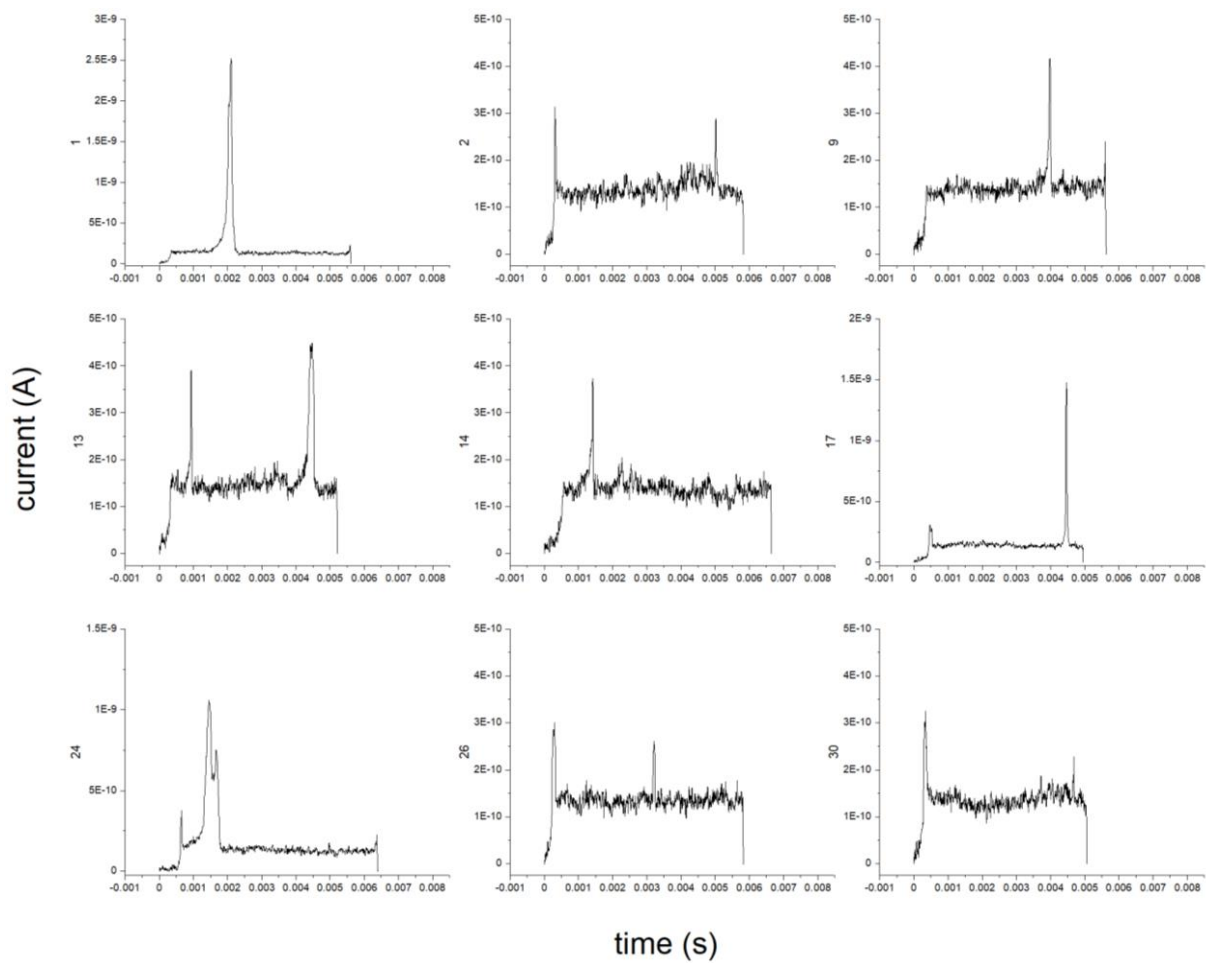

**Figure S12**: Current-time traces for the first 9 events formally identified with **two** sub-events. The current range is from -25 pA to 500 pA for all events, except events "1", "17" and "24", which have been re-scaled to an upper limit of 3 nA, 2 nA and 1.5 nA to improve visibility across the set.

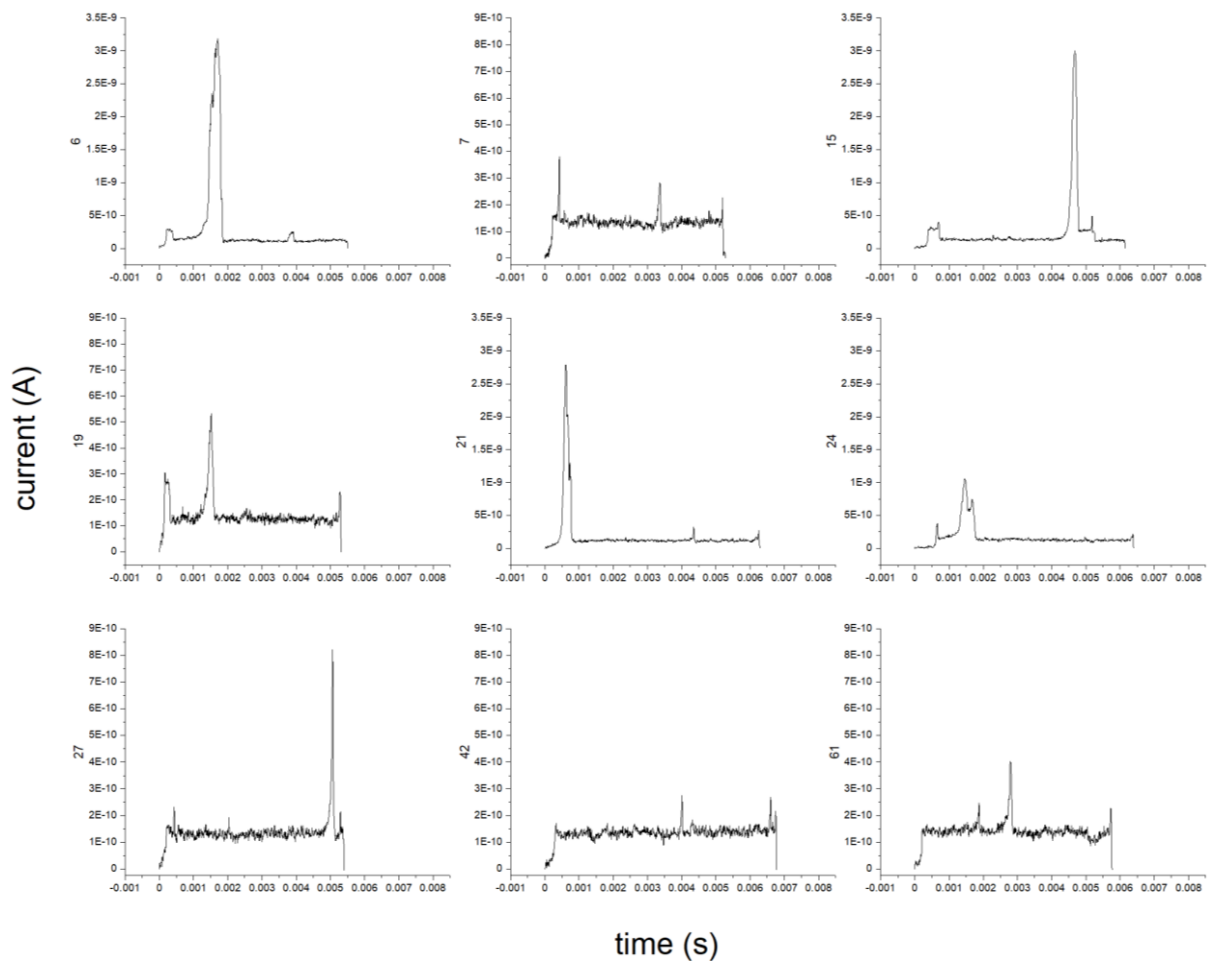

**Figure S13**: Current-time traces for the first 9 events formally identified with **three** sub-events. The current range is from -25 pA to 3.5 nA (events 6, 15, 21, 24) or -25 pA to 0.9 nA (all others).

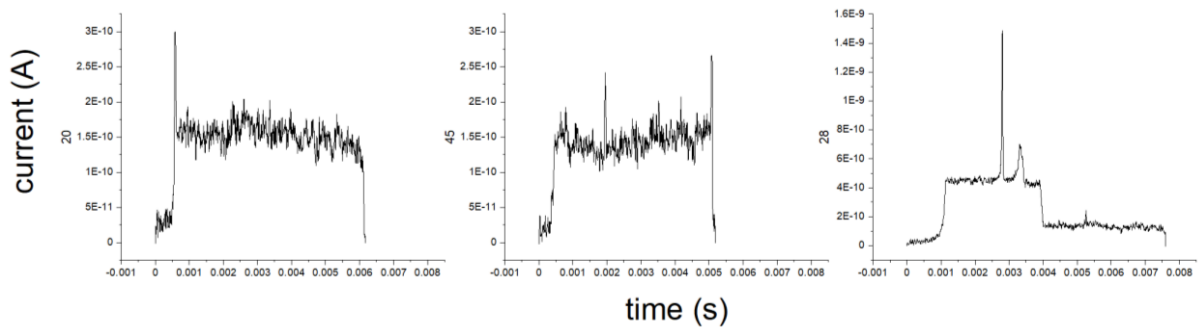

**Figure S14**: Current-time traces for three events that were formally identified with **4 or more** sub-events, the current range is from -25 pA to 325 pA (events 20 and 45, formally 4 sub-events) and from -100 pA to 1.6 nA (event 28, formally 12 sub-events). From visual inspection, these appear to be at least in part misclassifications, for example due to the underestimation of the event baseline level (events 20, 45) or the complex shape of the event (event 28).

**References:**

1. I. Teraoka, "Polymer Solutions - An Introduction to Physical Properties", Wiley & Sons (2002), chapter 3, p. 186